\documentclass[english,prl,twocolumn,superscriptaddress]{revtex4}
\usepackage{times}
\usepackage[T1]{fontenc}
\usepackage[latin1]{inputenc}
\usepackage{amsmath}
\usepackage{color}
\usepackage{graphicx}
\usepackage{amssymb}
\makeatletter
\usepackage{babel}
\makeatother

\begin{document}

\title{Field Tuned Superconductor to Insulator Transitions in an Amorphous Film with an Imposed Multiply Connected Geometry}

\date{\today{}}

\author{M. D. Stewart, Jr.}
\affiliation{Dept. of Physics, Brown University, Providence, RI 02912}
\author{Aijun Yin}
\affiliation{Division of Engineering, Brown University, Providence, RI 02912}
\author{J. M. Xu}
\affiliation{Dept. of Physics, Brown University, Providence, RI 02912}
\affiliation{Division of Engineering, Brown University, Providence, RI 02912}
\author{J. M. Valles, Jr.}
\affiliation{Dept. of Physics, Brown University, Providence, RI 02912}

\begin{abstract}
We have observed multiple magnetic field driven superconductor to insulator transitions (SIT) in amorphous Bi films perforated with a nano-honeycomb (NHC) array of holes. The period of the magneto-resistance, $H=H_M=h/2eS$ where $S$ is the area of a unit cell of holes, indicates the field driven transitions are boson dominated. The field-dependent resistance follows $R(T)=R_0(H)e^{T_0(H)/T}$ on both sides of the transition so that the evolution between these states is controlled by the vanishing of $T_0\to0$. We compare our results to the thickness driven transition in NHC films and the field driven transitions in unpatterned Bi films, other materials, and Josephson junction arrays. Our results suggest a structural source for similar behavior found in some materials and that despite the clear bosonic nature of the SITs, quasiparticle degrees of freedom likely also play an important part in the evolution of the SIT.
\end{abstract}
\maketitle
Quasi two-dimensional electronic systems exhibit a wide range of phenomena including the quantum Hall effect \cite{Hall}, weak and strong localization \cite{local}, and metal-insulator transitions \cite{Mott} including the superconductor to insulator transition (SIT) \cite{HebardPaalanen,Goldman-SITs,Wu-parallel}. These result from the enhancement of disorder, coulomb repulsion, and fluctuation effects on localization and superconductivity brought on by the reduced dimensionality. The aspects of these phenomena that are universal, or independent of the material comprising the systems, such as the exact quantization of the resistance plateaus in the quantum hall effect \cite{Hall} and the logarithmic decrease in the conductance of metal films with temperature \cite{Mott}, have made them particularly intriguing. 

Despite many experiments, a universal description \cite{Fisher-Bdrivensit} for the SIT in amorphous films, where a superconductor is driven to an insulating state by increasing disorder or magnetic field, has not been established. While many experiments show scaling behavior around a critical point near the quantum of resistance for Cooper pairs, $R_Q=h/4e^2$, qualitative differences have emerged between the SITs of different systems. For instance, amorphous InOx and TiN films \cite{Gantmakher-InOx,Shahar-scinsulator,Baturina-metalandpeakTiN} show a spectacular evolution from a superconducting to an exponentially insulating state with increasing magnetic field. Beyond this field tuned SIT, the magneto-resistance peaks at a value far in excess of both the normal state resistance, $R_N$, and $R_Q=h/4e^2$. Interestingly, not all amorphous materials show such dramatic behavior. In particular, amorphous MoGe \cite{MasonKapitulnik,Kapitulnik-quantumdissipation}, Ta \cite{Yoon-Ta}, Bi \cite{Goldman-a-Bi,Valles-QVL}, and Pb \cite{Xiong-miandbsit} show a resistance with a much weaker temperature dependence in their field induced insulating states and, at most, a very modest magneto-resistance peak \cite{Jay}. Some even exhibit a magnetic field induced metallic state \cite{MasonKapitulnik,Kapitulnik-quantumdissipation,Yoon-Ta,Valles-QVL,Okuma-MoSi}. While no complete explanation has taken root for this dichotomy, the most prominent explanations suggest that the spectacular behavior belies an underlying multiply connected geometry \cite{BaturinaVinokur,Meir-theoryofMRpeak}. 

We have taken an amorphous film system (Bi) which exhibits the latter, weaker behavior and transformed it into a multiply connected system by patterning it with a regular nano honeycomb (NHC) array of holes. This transformation results in a large magneto-resistance peak and a magnetic field induced exponentially localized insulating phase, the salient features of the data on InOx and TiN. Moreover, these films exhibit multiple SIT's with increasing field implying that Cooper pairs dominate both their superconducting and insulating ground states.

\begin{figure}
\begin{center}
\includegraphics[width=0.9\columnwidth,keepaspectratio]{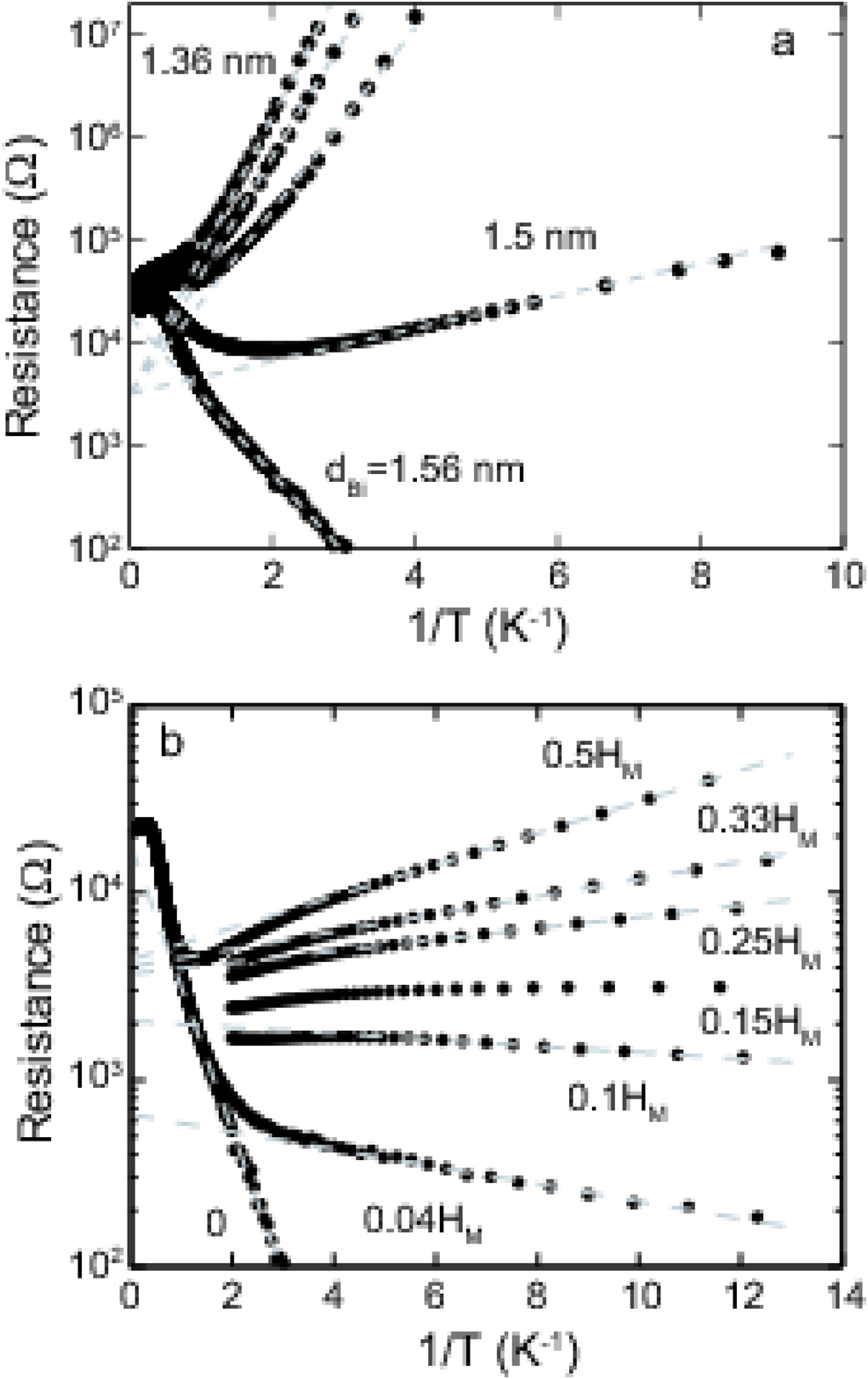}
\caption{a) Resistance versus inverse temperature for the thickness tuned superconductor to insulator transition. Dashed lines are fits to $R(T)=R_0e^{T_0/T}$. $T_0$ from most insulating to superconducting are 3.1, 2.65, 2.05, 0.37, and -1.76 K. b). Inverse temperature plot of the field driven IST bearing a strong resemblance to (a). Data in fields between $H=0$ and $H=H_M/2$ are nearly temperature independent. All dashed lines are fits to the same form with $T_0$s from top to bottom of 0.194, 0.108, 0.07, unfit, -0.04, -0.107, and -1.76 K. 
\label{cap:fig1}}
\end{center}
\end{figure}

These NHC films are produced within the ultra-high vacuum environment of our dilution refrigerator cryostat. The templates used are aluminum oxide substrates specially anodized to form the hole array \cite{XumakingAAOsubstrates}. Those substrates used in this study had a center to center spacing $a\simeq100$ nm and a radius $r_{hole}\simeq23$ nm. AFM topographs show the substrate surface regularly undulates with a period $\sim a$, so that NHC films contain a regular ($\sim 20\%$) variation in the film thickness. Before mounting in our cryostat these substrates are precoated with 10 nm of Ge and Au/Ge contact pads. Once at low temperature in the UHV environment of our dilution refrigerator, the substrates are coated with $<1$ nm of Sb to ensure the subsequent Bi films' amorphous morphology. Through sequential low-temperature evaporations of Bi we drive the system through a thickness tuned insulator to superconductor transition without breaking vacuum or warming (see Fig. \ref{cap:fig1}a).

The four-point transport measurements were made using standard low-frequency ac techniques and confirmed through dc measurements of the IV characteristics with battery operated electronics. The resistances of neighboring squares of the film were uniform to $5\%$. The magnetic field is supplied with a superconducting solenoid and is always perpendicular to the plane of the film.

The main data presented come from the superconducting film closest to the critical point of the disorder tuned SIT of a series of films (see Fig. \ref{cap:fig1}a). It has a normal state sheet resistance, $R_N$, of 21 k$\Omega$ and a mean field transition temperature, $T_{c0}\simeq1.5$ K \cite{thicknessdriven}. Below $T_{c0}$, the resistance, $R(T)$, drops exponentially with an activated form. The thinner insulating films also exhibit activated resistances with an activation energy of the opposite sign. Thus the superconducting state emerges from exponentially localized state and the transition can be characterized by a continuously varying energy scale \cite{Valles-critampfluc}.

The initial magnetic field driven SIT appears in Fig. \ref{cap:fig1}b. As the field is increased, the low temperature $dR/dT$ changes from superconducting-like, $dR/dT>0$ in $H=0$, to insulating-like, $dR/dT<0$ for $H>0.15H_M$. The low temperature resistance continues to increase up to $H_M/2$. Here, $H_M=h/2eS$ is called the matching field and $S$ is the area of a unit cell of holes. Figure \ref{cap:fig1}b shows the resistance can be parametrized by $R(T)=R_0(H)e^{T_0(H)/T}$ at the lowest temperatures in all fields except for the that in $H=0.15H_M$. The energy scale, $T_0$, extracted from some of the fits shown is small compared to the ultimate temperature and so cannot be associated with a simple activated process. Nonetheless, the superconducting and most insulating ($H_M/2$) data are well described by this form and it seems the most reasonable procedure for interpolating between them. 

\begin{figure}
\begin{center}
\includegraphics[width=0.9\columnwidth,keepaspectratio]{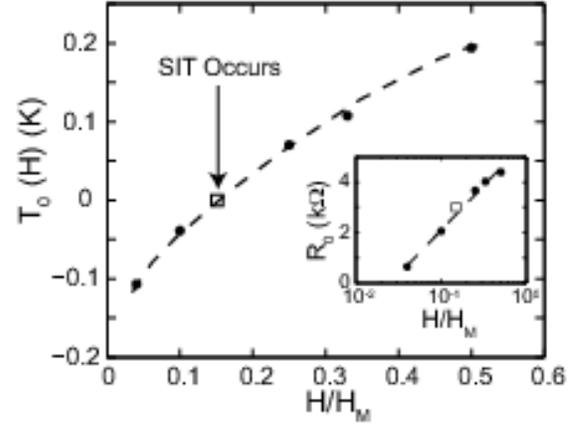}
\caption{Activation energies as a function of magnetic field which show the IST is governed by a single parameter $T_0\to0$. The box at $T_0(H)=0$ represents the data taken in $H=0.15H_M$ which is temperature independent. The dashed line is a guide to the eye. Inset shows $R_0(H)$ depends approximately logarithmically on field across the SIT. The dashed line in the inset is a fit to this functional form.
\label{cap:fig2}}
\end{center}
\end{figure}

The energy scale characterizing the transition, $T_0$, varies smoothly through the SIT and changes sign for $H\simeq0.15H_M$ (see Fig. \ref{cap:fig2}). Thus it appears $T_0\to0$ sets the critical field, $H_c$, while the value of the critical resistance is set by the other fit parameter, $R_0(H=.15H_M)\approx3$ k$\Omega$. The prefactor $R_0$ varies roughly logarithmically through the SIT (see Fig. \ref{cap:fig2} inset).

\begin{figure}
\begin{center}
\includegraphics[width=0.9\columnwidth,keepaspectratio]{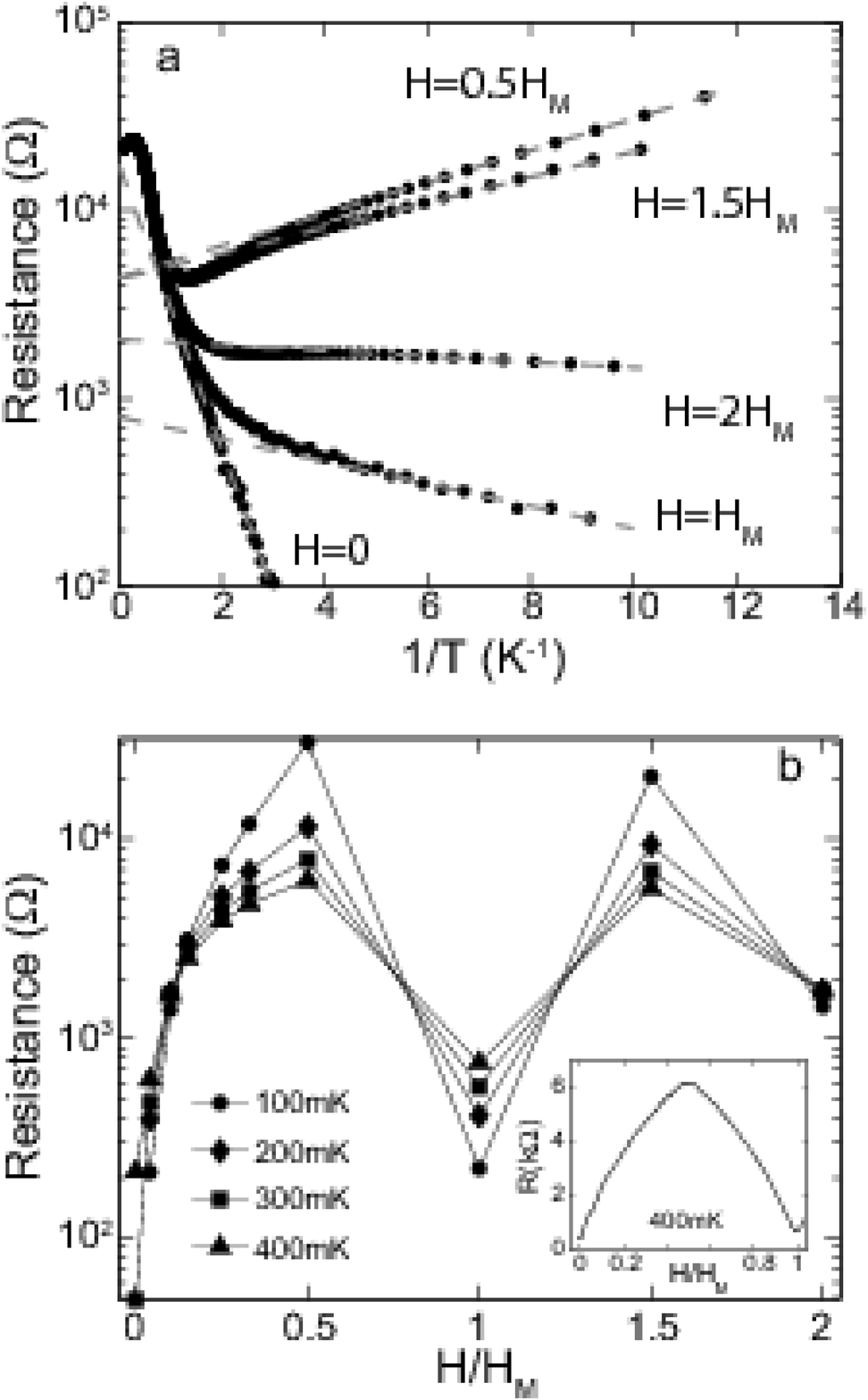}
\caption{a) $R(T)$ for half-matching and matching fields. Those data in matching fields show the tendency to superconduct as $T\to0$ showing there are at least three transitions between superconducting ($dR/dT>0$) and insulating ($dR/dT<0$) behavior with increasing field in the NHC film. Dashed lines are fits to $R(T)=R_0e^{(T_0/T)}$ with $T_0$ from top to bottom of 0.194, 0.157, -0.03, -0.138, and -1.76 K b) Magneto-resistance isotherms at a variety of temperatures. Inset shows a continuous field sweep over one period on a linear resistance scale. 
\label{cap:fig3}}
\end{center}
\end{figure}

Beyond $H_M/2$ $dR/dT$ and $T_0$ changes sign multiple times suggesting multiple field tuned SITs. Figure \ref{cap:fig3}a shows that $dR/dT>0$ at the matching fields 0, $H_M$, and $2H_M$ and $dR/dT<0$ at $H_M/2$, and $3H_M/2$ as appropriate for superconducting and insulating behaviors, respectively. The magneto-resistance isotherms in Fig. \ref{cap:fig2}a further illustrate this multiplicity of transitions and the underlying matching field period. The inset of Fig. \ref{cap:fig2}b provides the detailed shape of the first peak in the magneto-resistance. It appears symmetric about $H_M/2$. It is also apparent in Fig. \ref{cap:fig3}a that the magneto-resistance oscillations decrease in amplitude with increasing field. The ``superconducting'' state at $H_M$ is more resistive than at $H=0$ and the insulating state at $3H_M/2$ is less resistive than at $H_M/2$. These temperature dependences weaken further for subsequent SITs. In a field of a few Tesla, the associated magneto-resistance oscillations damp out completely.

\begin{figure}
\includegraphics[width=0.9\columnwidth,keepaspectratio]{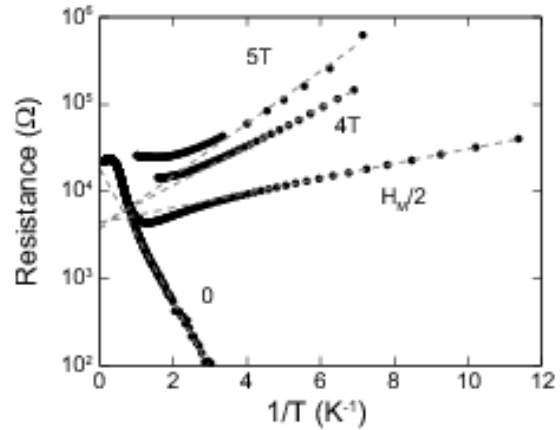}
\caption{Resistance behavior in high magnetic field for the $H=0$ superconducting film. At high fields the resistance is activated and the magnetoresistance positive suggesting the same transport mechanisms even when oscillations are small. $T_0$'s from top to bottom are 0.69, 0.51, 0.194, -1.76 K. 
\label{cap:fig4}}
\end{figure}

Beyond the oscillations, the films become increasingly insulating with magnetic field. Figure \ref{cap:fig4} also shows that the $R(T)$ in higher fields maintain an activated form (at least up to $H=4$ T) and the reentrance persists. Remarkably, the fit prefactor remains unchanged in a 4 T field from the value attained in $H_M/2$, nearly 20 times lower in field. 

The periodic appearance of ``superconducting'' states at matching fields set by the charge $2e$ flux quantum implies that Cooper pairs are the primary charge carriers throughout the field driven SIT. Thus, bosonic degrees of freedom dominate this transition. Moreover, comparison of the disorder and field tuned transitions suggest that similar bosonic physics operates in the two cases. There are three common features. Each exhibits activated resistances on both sides of the SIT and quasi-reentrant minima in the $R(T)$. Moreover, each can be characterized by a vanishing energy scale, $T_0$, at the critical point implying that the system is metallic there. These similarities suggest these SITs involve common mechanisms and that the magnetic field induces fluctuations of the same character as those induced by decreasing the thickness of NHC films.

The disorder and field tuned SITs of similar unpatterned films \cite{Xiong-miandbsit,Valles-critampfluc} differ qualitatively from the SITs of NHC films in a manner suggesting that fermionic, rather than bosonic degrees of freedom dominate their behavior. The $R(T)$ of unpatterned insulating films increase logarithmically rather than exponentially with decreasing temperature and they exhibit little, if any, reentrant behavior. Moreover, $H_c$ for unpatterned films approaches the expected upper critical field, $H_{c2}$, or the field at which the amplitude of the order parameter goes to zero. By contrast, $H_c\ll H_{c2}$ for NHC films, which allows multiple field tuned SITs. The above unpatterned film characteristics suggest that single electron motion dominates their insulating transport and order parameter amplitude fluctuations and pair breaking are strong in their superconducting phase. Thus, fermionic degrees of freedom influence both sides of their SIT.

On the other hand, the NHC films SITs qualitatively resemble those of Josephson Junction arrays (JJA). These lithographically prepared systems also show multiple SITs and $2e$ periodic magneto-resistance oscillations provided the ratio of the Josephson coupling, $E_J$, to the charging energy, $E_C$, is nearly critical (i.e. $E_J/E_C\simeq1$) \cite{VanderZant-JJarrays,Ootuka-shuntedJJarray}. JJAs also exhibit minima in their $R(T)$ and activated resistances \cite{VanderZant-JJarrays,haviland-cps} as do NHC films. While SITs do not occur at fractional values of $H/H_M$ for NHC films as they do in JJAs, it is likely that quenched disorder in the NHC lattice smears these out. The SITs of JJAs are usually discussed in terms of the bose condensation of vortices \cite{Fisher-Bdrivensit}. This picture seems to account for the $R(T)$ over a limited temperature range \cite{VanderZant-JJarrays}. It fails, however, for the lowest temperature data as the $R(T)$ become temperature independent in the $T\to0$ limit \cite{VanderZant-JJarrays}. This effect seems to require coupling to quasiparticle degrees of freedom in the array whether intentionally introduced through shunt resistors \cite{Ootuka-shuntedJJarray} or not \cite{VanderZant-JJarrays}. 

While the similarities above suggest that JJA physics drives the field driven SIT in the NHC films, there are some challenges to a direct comparison. First, the NHC films do not contain high quality tunnel junctions. All portions of NHC films are continuous, strongly disordered films without any high quality oxides. Second, the estimated charging energy using the nodes as the superconducting islands is far too large to compete with $E_J$ \cite{thicknessdriven}. Third, the $R(T)$ do not scale well around the critical point of the SIT as they do for the SITs of JJAs. We suggest that a model including quasi-particle degrees of freedom in addition to Bose degrees of freedom may be more appropriate for the NHC films \cite{Meir-theoryofMRpeak}. Virtual quasi-particle processes can screen Coulomb interactions to renormalize the charging energy to a value lower than the geometry implies \cite{Ootuka-shuntedJJarray,Ferrell}. Moreover, quasi-particle motion can provide a dissipative channel that could lead to metallic behavior and deviations from scaling \cite{MasonKapitulnik}.

The discussion above strongly suggests that patterning amorphous films into a multiply connected geometry promotes the influence of bosonic degrees of freedom. Recently, in fact, two models invoking such a geometry have emerged to explain the activated resistances and giant magneto-resistance peak near the SIT of InOx and TiN \cite{steiner-compareInOxandhighTc,Shahar-scinsulator,Baturina-metalandpeakTiN} films. In one picture \cite{Meir-theoryofMRpeak}, Cooper pair transport between islands \cite{Trivedi-inhomogeneouspairs,Ovadyahu-disordergrains} takes place in parallel with quasiparticle transport in the normal portions of the film. For $H<H_{peak}$, the magnetic field reduces phase coherence between islands, increasing the resistance. Quasiparticle transport is dominant for $H>H_{peak}$ as the field decreases the number and size of islands so that a peak is produced when the two mechanisms balance. In the second \cite{BaturinaVinokur}, the film is treated as an array of SQUIDs. Here the peak is the result of the magnetic field tuning $E_J\to0$. The position of the peak is then determined by the effective area of a squid loop in much the same way as the position of the first peak in Fig. \ref{cap:fig3}b is determined by the area of a unit cell of holes. Applying these ideas to NHC films seems to require elements of both explanations. The position and periodicity of the magnetoresistance peak is best explained through $E_J(H)\propto\cos(H)$ while the nearly metallic behavior of the $R(T,H)$ ($0\leq H\leq H_M/2$, $H_M$, and $2H_M$) seems to require dissipation. In each case, the inherent order parameter modulation \cite{BaturinaVinokur,Meir-theoryofMRpeak} is probably enhanced in NHC films through their patterning and small thickness variations. NHC film behavior, therefore, seems to provide a bridge between that of MoGe \cite{MasonKapitulnik,Kapitulnik-quantumdissipation} and Ta \cite{Yoon-Ta} on the one hand and TiN \cite{Baturina-metalandpeakTiN} and InOx \cite{Shahar-scinsulator} on the other.

We have measured multiple field driven SITs in an amorphous Bi film patterned with a nano-honeycomb array of holes. Each SIT is boson dominated, however, a description in terms of bose condensation falls short. Here, a full explanation would seem to require Josephson physics and quasiparticle degrees of freedom in an effectively islanded system. The activated resistances \cite{Shahar-scinsulator} and the magneto-resistance peak \cite{Baturina-metalandpeakTiN,Shahar-scinsulator} that NHC films share with TiN and InOx and the nearly metallic behavior shared with MoGe \cite{MasonKapitulnik,Kapitulnik-quantumdissipation} and Ta \cite{Baturina-metalandpeakTiN} indicate that NHC films occupy a space between these two extremes. Moreover, it seems a dissipative coupling and the type and degree of islanding interpolates between them.

We acknowledge helpful conversations with D. Feldman, and N. Trivedi. This work was supported by the NSF through DMR-0203608 and DMR-0605797, and by the AFRL and the ONR.
\bibliographystyle{apsrev}
\bibliography{refs}

\end{document}